\documentclass [11pt,a4paper] {article}

\usepackage[cp1252]{inputenc}

\usepackage{amssymb}
\usepackage{amsmath}
\usepackage{amsfonts,amssymb}
\usepackage[dvips]{graphicx}
\usepackage{bbm}
\usepackage{enumerate}

\usepackage{amsthm}

\usepackage{cancel}

\DeclareMathAlphabet{\mathpzc}{OT1}{pzc}{m}{it}

\def\SmallColSep{\setlength{\arraycolsep}{1pt}}

\begin{document}

\title{Probabilities in the logic of quantum propositions}

\author{Arkady Bolotin\footnote{$Email: arkadyv@bgu.ac.il$\vspace{5pt}} \\ \textit{Ben-Gurion University of the Negev, Beersheba (Israel)}}

\maketitle

\begin{abstract}\noindent In quantum logic, i.e., within the structure of the Hilbert lattice imposed on all closed linear subspaces of a Hilbert space, the assignment of truth values to quantum propositions (i.e., experimentally verifiable propositions relating to a quantum system) is unambiguously determined by the state of the system. So, if only pure states of the system are considered, can a probability measure mapping the probability space for truth values to the unit interval be assigned to quantum propositions? In other words, is a probability concept contingent or emergent in the logic of quantum propositions? Until this question is answered, the cause of probabilities in quantum theory cannot be completely understood. In the present paper it is shown that the interaction of the quantum system with its environment causes the irreducible randomness in the relation between quantum propositions and truth values.\\

\end{abstract}

\section{Introduction}  

\noindent According to the general belief, quantum mechanics is indeterministic \cite{Colbeck}, which implies that there does not exist a definite value determining in advance the outcome of every experiment performed on a quantum system.\\

\noindent And yet, attempts to treat \textit{indeterministically} quantum propositions (i.e., ones whose truth values can be determined with experiments on a quantum system) encounter some difficult questions.\\

\noindent To be sure, consider a quantum proposition denoted as $P$ which is represented by a closed linear subspace $\mathcal{H}_P$ of a Hilbert space $\mathcal{H}$ associated with a quantum system. If the system is prepared in the pure state $|\Psi_P\rangle$ belonging to the subspace $\mathcal{H}_P$, then it is reasonable to assume that any valuation (i.e., an assignment of truth values, \textit{true} and \textit{false}, to propositional variables) should assign the value of true to the quantum proposition $P$. In symbols, this can be expressed as $P(|\Psi_P\rangle) \mapsto \text{true}$  (where $P(|\Psi_P\rangle)$ stands for ``$P$ in the state $|\Psi_P\rangle$''). Similarly, in case that the system is prepared in the pure state $|\Psi^{\perp}_P\rangle$ residing in the subspace $\mathcal{H}_P^{\perp} \subseteq \mathcal{H}$ orthogonal to $\mathcal{H}_P$, it is reasonable to assign the value of false to the proposition $P$; in symbols, $P(|\Psi^{\perp}_P\rangle) \mapsto \text{false}$ .\\

\noindent Now, suppose that the system is prepared in the pure state $|\Psi_Q\rangle$ lying in the closed linear subspace $\mathcal{H}_Q \subseteq \mathcal{H}$, which represents another quantum proposition $Q$ so that $Q(|\Psi_Q\rangle) \mapsto \text{true}$ . Furthermore, suppose that the subspace $\mathcal{H}_Q$ is \textit{neither orthogonal nor co-aligned with the subspace $\mathcal{H}_P$}, which means that $\mathcal{H}_Q \nsubseteq \mathcal{H}_P^{\perp}$ and $\mathcal{H}_Q \nsupseteq \mathcal{H}_P^{\perp}$ as well as $\mathcal{H}_Q \nsubseteq \mathcal{H}_P$ and $\mathcal{H}_Q \nsupseteq \mathcal{H}_P$.\\

\noindent Allowing that quantum propositions obey quantum logic of Birkhoff and von Neumann \cite{Birkhoff}, one infers that the conjunction of $Q$ and $P$, i.e., $C = Q \wedge P$, must be always false and so $C(|\Psi_Q\rangle) \mapsto \text{false}$. This entails $P(|\Psi_Q\rangle) \mapsto \text{false}$; hence, the set of events $\{P(|\Psi_Q\rangle) \text{ is true}\}$ must contain no elements. However, according to Born's rule, there are instances in which $P$ is found out to be true when the system is prepared in $|\Psi_Q\rangle$. So, the question is, how can \textit{a non-zero probability} be assigned to the empty set $\{P(|\Psi_Q\rangle) \text{ is true}\}$ without invalidating axioms of probability?\\

\noindent To avoid this question, one might assume -- different from quantum logic -- that in the state $|\Psi_Q\rangle$ the quantum proposition $P$ is \textit{meaningless}, i.e., without a truth value $\mathfrak{t} \!\in\! \{\text{true}, \text{false}\}$, which can be written down in symbols as $P(|\Psi_Q\rangle) \not{\!\!\mapsto} \;\mathfrak{t}$. But then another question arises, namely, how does one evaluate the probability measure $\mu(P(|\Psi_Q\rangle))$ when the argument $P(|\Psi_Q\rangle)$ has no truth value? Put differently, how does the state $|\Psi_Q\rangle$ assign a probability-value to the quantum proposition $P$ without assigning it a truth-value first?\\

\noindent Consequently, the central question in probabilistic reasoning concerning quantum propositions is, how do such propositions become associated with probabilities? In other words, what is random about quantum propositions? Clearly, the origin of probabilities in quantum physics won’t be fully understood until this question is answered.\\

\noindent The present paper shows that the interaction of the quantum system with its environment brings about the irreducible randomness in the relation between quantum propositions and truth values.\\

\section{Admissibility and the indefiniteness of valuation}  

\noindent Recall that a proposition is identified with the set of interpretations which make the proposition true \cite{Gottwald}. One can infer from this that the assignment of truth-values to a quantum proposition, say $P$, can be defined by \textit{a predicate} (or rule), namely,\smallskip

\begin{equation} \label{PRED} 
   P(|\Psi\rangle)
   =
   \Phi_P(|\Psi\rangle)
   \;\;\;\;  ,
\end{equation}
\smallskip

\noindent which means that all states of the quantum system $|\Psi\rangle$ satisfying the predicate $\Phi_P$ render $P$ true.\\

\noindent The predicate $\Phi_P$ can be made precise. Let the system be prepared in the state $|\Psi\rangle$ residing in the closed linear subspace $\mathcal{H}_{|\Psi\rangle}$  of the Hilbert space $\mathcal{H}$ and let $\mathcal{H}_P \subseteq \mathcal{H}$ be the closed linear subspace that represents the quantum proposition $P$. Then $P$ assumes the value of true or false in accordance with the following rule\smallskip

\begin{equation} \label{FORM} 
   \Phi_P
   \text{:}
   \;\,
   |\Psi\rangle
   \in
   \mathcal{H}_{|\Psi\rangle} \wedge \mathcal{H}_P
   \;\mapsto\;\,
   \mathfrak{t} \in \{\text{true}, \text{false}\}
   \;\;\;\;  ,
\end{equation}
\smallskip

\noindent where $\mathcal{H}_{|\Psi\rangle} \wedge \mathcal{H}_P$ denotes the lattice-theoretic meet of the subspaces $\mathcal{H}_{|\Psi\rangle}$ and $\mathcal{H}_P$.\\

\noindent To see how this rule works, consider the Hilbert lattice $\mathcal{L}(\mathcal{H})$ imposed on all closed linear subspaces of $\mathcal{H}$.\\

\noindent Recall that closed linear subspaces, say, $\mathcal{H}_A$ and $\mathcal{H}_B$, of a Hilbert space $\mathcal{H}$ are called \textit{commutable} if the following condition holds \cite{Pavicic}:\smallskip

\begin{equation} \label{COMM} 
   \mathcal{H}_A
   \cap
   \left(
      \mathcal{H}_A
      \cap
      \mathcal{H}_B^{\perp}
   \right)^{\perp}
   \subseteq
   \mathcal{H}_B
   \;\;\;\;  ,
\end{equation}
\smallskip

\noindent where $\cap$ denotes the set-theoretic intersection and $(\cdot)^{\perp}$ stands for the orthogonal complement of $(\cdot)$.\\

\noindent Suppose that $\mathcal{H}_{|\Psi\rangle}$  is co-aligned with $\mathcal{H}_P$ such that $\mathcal{H}_{|\Psi\rangle} \subseteq \mathcal{H}_P$. Then, $\mathcal{H}_{|\Psi\rangle}^{\perp} \supseteq \mathcal{H}_P^{\perp}$, and so $\mathcal{H}_{|\Psi\rangle} \cap \mathcal{H}_P^{\perp} = \{0\}$ which means that the condition (\ref{COMM}) holds: $\mathcal{H}_{|\Psi\rangle} \cap (\mathcal{H}_{|\Psi\rangle} \cap \mathcal{H}_P^{\perp})^{\perp} = \mathcal{H}_{|\Psi\rangle} \subseteq \mathcal{H}_P$. At the same time, because $\mathcal{H}_{|\Psi\rangle} \wedge \mathcal{H}_P = (\mathcal{H}_{|\Psi\rangle} \cap \mathcal{H}_P) = \mathcal{H}_{|\Psi\rangle}$, in place of $|\Psi\rangle \in \mathcal{H}_{|\Psi\rangle} \wedge \mathcal{H}_P$ one finds the expression $|\Psi\rangle \in \mathcal{H}_{|\Psi\rangle}$ which is true under the initial condition. Hence,\smallskip

\begin{equation}  
   \Phi_P
   \text{:}
   \;\,
   |\Psi\rangle
   \in
   \mathcal{H}_{|\Psi\rangle}
   \;\mapsto\;\,
   \text{true}
   \;\;\;\;  ,
\end{equation}
\smallskip

\noindent and, in accordance with (\ref{PRED}), the quantum proposition $P$ assumes the value of true in $|\Psi\rangle \in \mathcal{H}_P$.\\

\noindent Next, suppose that $\mathcal{H}_{|\Psi\rangle}$  is orthogonal to $\mathcal{H}_P$ and so $\mathcal{H}_{|\Psi\rangle} \cap \mathcal{H}_P = \{0\}$. Let $\mathcal{H}_{|\Psi\rangle} \subseteq \mathcal{H}_P^{\perp}$; this means $(\mathcal{H}_{|\Psi\rangle} \cap \mathcal{H}_P^{\perp}) = \mathcal{H}_{|\Psi\rangle}$, so the condition (\ref{COMM}) is valid: $\mathcal{H}_{|\Psi\rangle} \cap (\mathcal{H}_{|\Psi\rangle} \cap \mathcal{H}_P^{\perp})^{\perp} = \{0\} \subseteq \mathcal{H}_P$. If $\mathcal{H}_{|\Psi\rangle} \supseteq \mathcal{H}_P^{\perp}$, this condition is valid again: $\mathcal{H}_P \cap (\mathcal{H}_P \cap \mathcal{H}_{|\Psi\rangle}^{\perp})^{\perp} = \{0\} \subseteq \mathcal{H}_{|\Psi\rangle}$. On the other hand, because any (physically meaningful) state $|\Psi\rangle$ differs from 0, it follows\smallskip

\begin{equation}  
   \Phi_P
   \text{:}
   \;\,
   |\Psi\rangle
   \in
   \mathcal{H}_{|\Psi\rangle} \wedge \mathcal{H}_P = \{0\}
   \;\mapsto\;\,
   \text{false}
   \;\;\;\;  ,
\end{equation}
\smallskip

\noindent meaning that the quantum proposition $P$ takes on the value of false in $|\Psi\rangle \in \mathcal{H}^{\perp}_P$.\\

\noindent Let $\mathcal{H}_{|\Psi\rangle} \supseteq \mathcal{H}_P$; so therefore, the condition (\ref{COMM}) holds once again: $\mathcal{H}_P \cap (\mathcal{H}_P \cap \mathcal{H}_{|\Psi\rangle}^{\perp})^{\perp} \subseteq \mathcal{H}_{|\Psi\rangle}$. Using (\ref{FORM}) one finds that\smallskip

\begin{equation}  
   \Phi_P
   \text{:}
   \;\,
   |\Psi\rangle
   \in
   \left(
      \mathcal{H}_{|\Psi\rangle} \cap \mathcal{H}_P
   \right)
   =
   \mathcal{H}_P
   \;\mapsto\;\,
   \left\{
      \!\!
      \begin{array}{l}
         \text{true}
         ,\;\;\;
         |\Psi\rangle \in \mathcal{H}_P
         \\ 
         \text{false}
         ,\;\;\;
         |\Psi\rangle \notin \mathcal{H}_P
      \end{array}
   \right.
   \;\;\;\;  ,
\end{equation}
\smallskip

\noindent i.e., the quantum proposition $P$ is true or false depending on whether or not the state $|\Psi\rangle$ lying in $\mathcal{H}_{|\Psi\rangle}$ belongs to $\mathcal{H}_P$ as well.\\

\noindent Now, suppose that the closed linear subspaces $\mathcal{H}_{|\Psi\rangle}$ and $\mathcal{H}_P$ do not commute, and, as a result, they are neither orthogonal nor co-aligned. Notwithstanding this stipulation, the meet $\mathcal{H}_{|\Psi\rangle} \wedge \mathcal{H}_P$ exists even then because the Hilbert lattice $\mathcal{L}(\mathcal{H})$ is \textit{complete} \cite{Ptak, Redei}. What is more, in the lattice $\mathcal{L}(\mathcal{H})$ this meet corresponds to the zero-dimensional subspace $\{0\}$. Consequently, one obtains that\smallskip

\begin{equation}  
   \Phi_P
   \text{:}
   \;\,
   |\Psi\rangle
   \in
   \left(
      \mathcal{H}_{|\Psi\rangle} \wedge \mathcal{H}_P
   \right)
   =
   \{0\}
   \;\mapsto\;\,
   \text{false}
   \;\;\;\;  ,
\end{equation}
\smallskip

\noindent i.e., the quantum proposition $P$ assumes the value of false in the said case.\\

\noindent It follows from this analysis that within the structure of the Hilbert lattice $\mathcal{L}(\mathcal{H})$ quantum propositions can only be \textit{truth-value definite}, which leaves no room for a random phenomenon.\\

\noindent A way to bring indefiniteness into a valuation of quantum propositions is \textit{to strengthen the rule} (\ref{FORM}) by the additional requirement of \textit{admissibility}. Specifically, one may demand that the predicate $\Phi_P$ will return a truth value only if $\Phi_P$ is admissible in the state $|\Psi\rangle$; if it is not, then the quantum proposition $P$ should remain truth-value indefinite in $|\Psi\rangle$.\\

\noindent To formulate conditions of admissibility some preliminaries are in order first.\\

\noindent Recall that a closed linear subspace, say $\mathcal{H}_P$, of a Hilbert space $\mathcal{H}$ can be considered as \textit{a range} of some projection operator (i.e., self-adjoint idempotent operator), say $\hat{P}$, acting on $\mathcal{H}$. Explicitly, the subspace $\mathcal{H}_P$ is identical to the subset of the vectors $|\Psi\rangle \in \mathcal{H}$ that are in the image of $\hat{P}$:\smallskip

\begin{equation} \label{RAN} 
   \mathcal{H}_P
   \equiv
   \mathrm{ran}(\hat{P})
   =
   \left\{
      |\Psi\rangle \in \mathcal{H}
      \text{:}
      \;
      \hat{P} |\Psi\rangle =  |\Psi\rangle
   \right\}
   \;\;\;\;  .
\end{equation}
\smallskip

\noindent In the same way, the closed linear subspace orthogonal to $\mathcal{H}_P$ is identical to \textit{the kernel} of $\hat{P}$, i.e., the subset of the vectors $|\Psi\rangle \in \mathcal{H}$ that are mapped to zero by $\hat{P}$, explicitly:\smallskip

\begin{equation} \label{KER} 
   \mathcal{H}_P^{\perp}
   \equiv
   \mathrm{ker}(\hat{P})
   =
   \mathrm{ran}(\hat{1} - \hat{P})
   =
   \left\{
      |\Psi\rangle \in \mathcal{H}
      \text{:}
      \;
      (\hat{1} - \hat{P}) |\Psi\rangle =  |\Psi\rangle
   \right\}
   \;\;\;\;  ,
\end{equation}
\smallskip

\noindent where $\hat{1}$ stands for the identity operator on $\mathcal{H}$. For that reason, the projection operator $\hat{1}-\hat{P}$ can be understood as the negation of $\hat{P}$, i.e.,\smallskip

\begin{equation}  
   \neg\hat{P}
   =
   \hat{1} - \hat{P}
   \;\;\;\;  .
\end{equation}
\smallskip

\noindent As consequences of (\ref{RAN}) and (\ref{KER}), one has\smallskip

\begin{equation}  
   \mathrm{ran}(\hat{P})
   \cap
   \mathrm{ran}(\neg\hat{P})
   =
   \mathrm{ran}(\hat{0})
   =
   \{0\}
   \;\;\;\;  ,
\end{equation}

\begin{equation}  
   \mathrm{ran}(\hat{P})
   +
   \mathrm{ran}(\neg\hat{P})
   =
   \mathrm{ran}(\hat{1})
   =
   \mathcal{H}
   \;\;\;\;  ,
\end{equation}
\smallskip

\noindent where $\hat{0}$ is the zero operator on $\mathcal{H}$, while the subsets $\{0\}$ and $\mathcal{H}$ are \textit{the trivial subspaces of $\mathcal{H}$} (which correspond to the trivial projection operators $\hat{0}$ and $\hat{1}$, respectively).\\

\noindent Recall that the set of two or more nontrivial projection operators $\hat{P}_A$, $\hat{P}_B$, … on $\mathcal{H}$ is called \textit{a context} $\Sigma$\smallskip

\begin{equation}  
   \Sigma
   =
   \left\{
      \hat{P}_A, \hat{P}_B, \dots
   \right\}
   \;\;\;\;   
\end{equation}
\smallskip

\noindent if the next requirements are satisfied:\smallskip

\begin{equation}  
   \hat{P}_A
   \hat{P}_B
   =
   \hat{P}_B
   \hat{P}_A
   =
   \hat{0}
   \;\;\;\;  ,
\end{equation}

\begin{equation}  
   \hat{P}_A
   +
   \hat{P}_B
   +
   \dots
   =
   \sum_{\hat{P} \in \Sigma}
   \hat{P}
   =
   \hat{1}
   \;\;\;\;  .
\end{equation}
\smallskip

\noindent Let $\mathcal{O} = \{\Sigma\}$ be the set of all the contexts associated with the quantum system and let $P_A$ and $P_B$ denote quantum propositions represented by $\mathrm{ran}(\hat{P}_A)$ and $\mathrm{ran}(\hat{P}_B)$ respectively. Then, the rule (\ref{FORM}) will be admissible for a valuation of $P_A$ and $P_B$ in $|\Psi\rangle$ if the following two conditions are applicable for any $\Sigma \in \mathcal{O}$ \cite{Abbott}:\smallskip

\renewcommand{\labelenumi}{(\roman{enumi}).}
\begin{enumerate}
\item 	if there is such $\hat{P}_A \in \Sigma$ that $P_A(|\Psi\rangle)$ assumes the value of true, then $P_B(|\Psi\rangle)$ must be false for all $\hat{P}_B \in (\Sigma \!\setminus\! \{\hat{P}_A\})$;
\item 	if there is such $\hat{P}_A \in \Sigma$ that $P_B(|\Psi\rangle)$ assumes the value of false for all $\hat{P}_B \in (\Sigma \!\setminus\! \{\hat{P}_A\})$, then $P_A(|\Psi\rangle)$ must be true.
\end{enumerate}

\noindent It is not difficult to see that the validity of the admissibility conditions relies on the mutual commutability of the subspaces $\mathcal{H}_{|\Psi\rangle}$ and $\mathcal{H}_P$. Thus, if the contexts $\Sigma_{|\Psi\rangle}$ and $\Sigma_P$ containing in that order the projection operators $|\Psi\rangle\langle\Psi|$ and $\hat{P}$ are \textit{non-intertwined} (meaning that $\Sigma_{|\Psi\rangle}$ and $\Sigma_P$ do not share common projection operators \cite{Svozil}), then the subspaces $\mathcal{H}_{|\Psi\rangle} = \mathrm{ran}(|\Psi\rangle\langle\Psi|)$ and $\mathcal{H}_P = \mathrm{ran}(\hat{P})$ will be incommutable. According to the rule (\ref{FORM}), this means that the quantum proposition $P$ should assume the value of false in the states $|\Psi\rangle \!\in\! \mathcal{H}_{|\Psi\rangle}$ for all $\hat{P} \!\in\! \Sigma_P$. This violates the condition ($\mathrm{ii}$); consequently, the rule (\ref{FORM}) should be considered inadmissible for a valuation of the quantum proposition $P$ in $|\Psi\rangle \!\in\! \mathcal{H}_{|\Psi\rangle}$, and so $P$ should remain truth-value indefinite in these states.\\

\noindent However, be that as it might, one must not forget that the admissibility requirement is in fact some extraneous supposition coming from outside the Hilbert space formalism. Another way to put it is that admissibility is not belonging to the structure of the Hilbert lattice $\mathcal{L}(\mathcal{H})$. Hence, the truth-value indefiniteness of quantum propositions introduced through admissibility is contingent (not necessary existing) and, for this reason, cannot be considered intrinsic.\\

\section{Indefinite valuation in supervaluationism}  

\noindent An alternative way to introduce indefiniteness into a valuation of quantum propositions is \textit{to not impose the structure of the Hilbert lattice $\mathcal{L}(\mathcal{H})$} on closed linear subspaces of a Hilbert space $\mathcal{H}$.\\

\noindent To give details of this approach, recall first that a subspace $\mathcal{H}_A \subseteq \mathcal{H}$ is called \textit{an invariant subspace} under the projection operator $\hat{P}_A$ on $\mathcal{H}$ if\smallskip

\begin{equation}  
   \hat{P}
   \text{:}
   \;
   \mathcal{H}_A
   \to
   \mathcal{H}_A
   \;\;\;\;  .
\end{equation}
\smallskip

\noindent This means that the image of every vector $|\Psi_A\rangle$ in $\mathcal{H}_A$ under $\hat{P}_A$ remains within $\mathcal{H}_A$ which can be denoted as\smallskip

\begin{equation}  
   \hat{P}_A \mathcal{H}_A
   =
   \left\{
      |\Psi_A\rangle \in \mathcal{H}_A
      \text{:}
      \;\;
      \hat{P}_A|\Psi_A\rangle
   \right\}
   \subseteq
   \mathcal{H}_A
   \;\;\;\;  .
\end{equation}
\smallskip

\noindent Let $\mathcal{L}(\hat{P}_A)$ refer to the set of the invariant subspaces of $\mathcal{H}$ invariant under the projection operator $\hat{P}_A$:\smallskip

\begin{equation}  
   \mathcal{L}(\hat{P}_A)
   =
   \left\{
      \mathcal{H}_A \subseteq \mathcal{H}
      \text{:}
      \;\;
      \hat{P}_A \mathcal{H}_A
      \subseteq
      \mathcal{H}_A
   \right\}
   \;\;\;\;  .
\end{equation}
\smallskip

\noindent Consider the set $\mathcal{L}(\Sigma)$ of the invariant subspaces which are invariant under every projection operator $\hat{P}_A$, $\hat{P}_B$, $\dots$ from the context $\Sigma$:\smallskip

\begin{equation}  
   \mathcal{L}(\Sigma)
   =
   \mathcal{L}(\hat{P}_A)
   \cap
   \mathcal{L}(\hat{P}_B)
   \cap
   \dots
   =
   \bigcap_{\hat{P} \in \Sigma}
   \mathcal{L}(\hat{P})
   \;\;\;\;  .
\end{equation}
\smallskip

\noindent The elements of this set form a complete lattice called \textit{the invariant-subspace lattice of the context $\Sigma$} \cite{Radjavi}. The lattice operations on $\mathcal{L}(\Sigma)$ are defined in an ordinary way: Specifically, the meet $\wedge$ and the join $\vee$ are defined by\smallskip

\begin{equation}  
   \mathcal{H}_A
   ,
   \mathcal{H}_B
   \in
   \mathcal{L}(\Sigma)
   \;\;
   \implies
   \;\;
   \left\{
      \begin{array}{l}
         \mathcal{H}_A
         \wedge
         \mathcal{H}_B
         =
         \mathcal{H}_A
         \cap
         \mathcal{H}_B
         \in
         \mathcal{L}(\Sigma)
         \\ 
         \mathcal{H}_A
         \vee
         \mathcal{H}_B
         =
         \left(
            (\mathcal{H}_A)^{\perp}
            \cap
            (\mathcal{H}_B)^{\perp}
         \right)^{\perp}
         \in
         \mathcal{L}(\Sigma)
      \end{array}
   \right.
   \;\;\;\;  .
\end{equation}
\smallskip

\noindent It is straightforward to verify that each invariant-subspace lattice $\mathcal{L}(\Sigma)$ contains only mutually commuting subspaces (corresponding to mutually commutable projection operators), which means that each $\mathcal{L}(\Sigma)$ is a Boolean algebra. It follows then that for all $\mathrm{ran}(\hat{P}_A), \mathrm{ran}(\hat{P}_B) \in \mathcal{L}(\Sigma)$, one has\smallskip

\begin{equation}  
      \mathrm{ran}(\hat{P}_A)
      \wedge
      \mathrm{ran}(\hat{P}_B)
      =
      \{0\}
      \;\;
      \iff
      \;\;
      \hat{P}_A
      \hat{P}_B
      =
      \hat{P}_B
      \hat{P}_A
      =
      \hat{0}
   \;\;\;\;  ,
\end{equation}

\begin{equation}  
      \mathrm{ran}(\hat{P}_A)
      \wedge
      \mathrm{ran}(\hat{P}_B)
      =
      \mathrm{ran}(\hat{P}_A)
      \;\;
      \iff
      \;\;
      \hat{P}_A
      \hat{P}_B
      =
      \hat{P}_B
      \hat{P}_A
      =
      \hat{P}_A
   \;\;\;\;  ,
\end{equation}

\begin{equation}  
      \mathrm{ran}(\hat{P}_A)
      +
      \mathrm{ran}(\hat{P}_B)
      +
      \dots
      =
      \mathcal{H}
      \;\;
      \iff
      \;\;
      \hat{P}_A
      +
      \hat{P}_B
      +
      \dots
      =
      \hat{1}
   \;\;\;\;  .
\end{equation}
\smallskip

\noindent Consider the collection of the lattices $\{\mathcal{L}(\Sigma)\}_{\Sigma \in \mathcal{O}}$ which is in one-to-one correspondence with the set of all the contexts $\mathcal{O}$ associated with the quantum system. Since $\hat{P}\{0\} \subseteq \{0\}$ and $\hat{P} \mathcal{H} \subseteq \mathcal{H}$ for all $\hat{P} \in \mathcal{O}$, the trivial subspaces $\{0\}$ and $\mathcal{H}$ are elements of each invariant-subspace lattice from $\{\mathcal{L}(\Sigma)\}_{\Sigma \in \mathcal{O}}$.\\

\noindent If all the lattices from $\{\mathcal{L}(\Sigma)\}_{\Sigma \in \mathcal{O}}$ are pasted (stitched) together at their common elements -- that is, in any case, at the trivial subspaces $\{0\}$ and $\mathcal{H}$ -- then the resulted logic will be the Hilbert lattice $\mathcal{L}(\mathcal{H})$. In this sense, the Hilbert lattice $\mathcal{L}(\mathcal{H})$ can be thought of as \textit{the pasting of the invariant-subspace lattices from the collection} $\{\mathcal{L}(\Sigma)\}_{\Sigma \in \mathcal{O}}$. Providing the set of all the contexts $\mathcal{O}$ form a continuum, the Hilbert lattice is a continuum of pasting of the Boolean algebras $\mathcal{L}(\Sigma) \in \{\mathcal{L}(\Sigma)\}_{\Sigma \in \mathcal{O}}$.\\

\noindent However, without such pasting construction, i.e., giving up the assumption of the Hilbert lattice, the structure of $\{\mathcal{L}(\Sigma)\}_{\Sigma \in \mathcal{O}}$ results in the logic that can be identified as \textit{supervaluationism}.\\

\noindent To be sure, recall that supervaluation semantics retains the classical consequence relation and classical laws at the same time as admitting \textit{truth-value gaps} \cite{Varzi, Keefe}. For example, according to supervaluationism, a disjunction $D = P_B \vee \neg P_B$ and a conjunction $C = P_B \wedge \neg P_B$ assume the value of true and false, respectively, even when $P_B$ and $\neg P_B$ have no truth values at all (i.e., have truth values gaps).\\

\noindent Assuming that the lattice-theoretic meet $\wedge$ and join $\vee$ of commuting subspaces can be interpreted as the conjunction and disjunction of the quantum propositions represented by those subspaces, one finds that $D$ and $C$ must be represented by the trivial subspaces $\mathcal{H}$ and $\{0\}$, respectively, because\smallskip

\begin{equation}  
      \mathrm{ran}(\hat{P}_B)
      +
      \mathrm{ran}(\neg\hat{P}_B)
      =
      \mathrm{ran}(\hat{P}_B)
      \vee
      \mathrm{ran}(\neg\hat{P}_B)
      =
      \mathcal{H}
   \;\;\;\;  ,
\end{equation}

\begin{equation}  
      \mathrm{ran}(\hat{P}_B)
      \wedge
      \mathrm{ran}(\neg\hat{P}_B)
      =
      \{0\}
   \;\;\;\;  ,
\end{equation}
\smallskip

\noindent where the subspaces $\mathrm{ran}(\hat{P}_B)$ and $\mathrm{ran}(\neg\hat{P}_B)$ represent the quantum proposition $P_B$ and its negation $\neg P_B$. Next, it follows that in any pure state of the system, for example, $|\Psi_A\rangle \in \mathrm{ran}(\hat{P}_A)$, the predicates $\Phi_D$ for $D$ and $\Phi_C$ for $C$ return the values of true and false correspondingly:\smallskip

\begin{equation}  
   \Phi_D
   \text{:}
   \;\,
   |\Psi_A\rangle
   \in
   \left(
      \mathrm{ran}(\hat{P}_A) \cap \mathcal{H}
   \right)
   =
   \mathrm{ran}(\hat{P}_A)
   \;\mapsto\;\,
   \text{true}
   \;\;\;\;  ,
\end{equation}

\begin{equation}  
   \Phi_C
   \text{:}
   \;\,
   |\Psi_A\rangle
   \in
   \left(
      \mathrm{ran}(\hat{P}_A) \cap \{0\}
   \right)
   =
   \{0\}
   \;\mapsto\;\,
   \text{false}
   \;\;\;\;  ,
\end{equation}
\smallskip

\noindent so, the disjunction $D$ is a tautology while the conjunction $C$ is a contradiction.\\

\noindent On the other hand, if the projection operators $\hat{P}_A$ and $\hat{P}_B$ belong to the non-intertwined contexts $\Sigma_A$ and $\Sigma_B$, respectively, neither $\mathrm{ran}(\hat{P}_B)$ nor $\mathrm{ran}(\neg\hat{P}_B)$ can be elements of the invariant-subspace lattice $\mathcal{L}(\Sigma_A)$ containing $\mathrm{ran}(\hat{P}_A)$. In the said case, not having the pasting construction of the Hilbert lattice $\mathcal{L}(\mathcal{H})$ means that the lattice-theoretic meet $\wedge$ cannot be defined on a pair of $\mathrm{ran}(\hat{P}_A)$ and $\mathrm{ran}(\hat{P}_B)$ as well as on a pair of $\mathrm{ran}(\hat{P}_A)$ and $\mathrm{ran}(\neg\hat{P}_B)$ (recall that the meet is defined as an operation on pairs of elements from one lattice \cite{Davey}). Hence, in the states $|\Psi_A\rangle \in \mathrm{ran}(\hat{P}_A)$, the predicates $\Phi_{P_B}$ for $P_B$ and $\Phi_{\neg P_B}$ for $\neg P_B$ cannot be determined and so return no truth value (without a supplementary requirement of admissibility). In symbols,\smallskip

\begin{equation}  
   \hat{P}_A
   ,
   \hat{P}_B
   \notin
   \left(
      \Sigma_A
      \cap
      \Sigma_B
   \right)
   \;\;
   \implies
   \;\;
   \Phi_{\beta}
   \text{:}
   \;\;
   |\Psi_A\rangle
   \in
   \mathrm{ran}(\hat{P}_A)
   \,\cancel{\;\wedge\;}\,
   \mathrm{ran}(\hat{\beta})
   \,
   \not{\!\!\mapsto}
   \;\;
   \mathfrak{t}
   \;\;\;\;  ,
\end{equation}
\smallskip

\noindent where $\beta \in \{ P_B, \neg P_B \}$, $\hat{\beta} \in \{ \hat{P}_B, \neg\hat{P}_B \}$, and the diagonal strikeout of $\wedge$ indicates that the meet operation cannot be defined. In consequence, the quantum proposition $P_B$ and its negation $\neg P_B$ are indefinite (i.e., ``gappy'') in the states $|\Psi_A\rangle \in \mathrm{ran}(\hat{P}_A)$, although their disjunction $D$ and conjunction $C$ have truth values in these states.\\

\noindent As one can see, the aforesaid truth-value indefiniteness is proper to the structure of $\{\mathcal{L}(\Sigma)\}_{\Sigma \in \mathcal{O}}$ and accordingly can be regarded as intrinsic.\\

\noindent To elucidate how ``gappy'' quantum propositions become associated with probabilities in $\{\mathcal{L}(\Sigma)\}_{\Sigma \in \mathcal{O}}$, consider the following simple model.\\

\section{A two-state model for probabilities}  

\noindent Let a qubit -- i.e., a two-state quantum-mechanical system (such as a one-half spin particle, say, an electron) -- which is denoted as $S$, interact with its environment $E$ described by a collection of $N$ other qubits.\\

\noindent Suppose that each environmental qubit has the preferred set of states (say, due to the design of the experiment) corresponding to the eigenvalues $+1$ and $-$1 of the Pauli matrix $\sigma_z$, namely,\smallskip

\begin{equation}  
   \left\{
      |\Psi_{kz+}\rangle
      ,
      |\Psi_{kz-}\rangle
   \right\}
   \equiv
   \left\{
         \left[
            \begingroup\SmallColSep
            \begin{array}{r}
               1
               \\
               0
            \end{array}
            \endgroup
         \right]
       ,
         \left[
            \begingroup\SmallColSep
            \begin{array}{r}
               0
               \\
               1
            \end{array}
            \endgroup
         \right]
   \right\}
   \;\;\;\;  ,
\end{equation}
\smallskip

\noindent for all $k \in \{1,\dots,N\}$. Correspondingly, the Hilbert space of the $k^{\mathrm{th}}$  environmental qubit $\mathcal{H}_k$ is\smallskip

\begin{equation}  
   \mathcal{H}_k
   =
   \mathrm{ran}(\hat{1})
   =
   \sum_{\beta = \pm}
   \mathrm{ran}(\hat{P}_{kz\beta})
   \;\;\;\;  ,
\end{equation}
\smallskip

\noindent where $\hat{P}_{kz\pm}$ are the projection operators of spin along the $z$-axis.\\

\noindent Also suppose that, in contrast to the environmental qubits, the qubit $S$ has no preferred states and so the Hilbert space $\mathcal{H}_S$ of the qubit $S$ can be presented as\smallskip

\begin{equation}  
   \mathcal{H}_S
   =
   \mathrm{ran}(\hat{1})
   =
   \sum_{\alpha = \pm}
   \mathrm{ran}(\hat{P}_{Su^{\prime}\alpha})
   =
   \sum_{\alpha = \pm}
   \mathrm{ran}(\hat{P}_{Su^{\prime\prime}\alpha})
   =
   \dots
   \;\;\;\;  ,
\end{equation}
\smallskip

\noindent where $u^{\prime}, u^{\prime\prime}, \dots \in \mathbb{R}^3$ are arbitrary axes.\\ 

\noindent Since for all $u^{\prime} \!\neq u^{\prime\prime}$ the projection operators $\hat{P}_{Su^{\prime}\pm}$ and $\hat{P}_{Su^{\prime\prime}\pm}$ belong to the different contexts, namely, $\Sigma_{Su^{\prime}}$ and $\Sigma_{Su^{\prime\prime}}$, the subspaces $\mathrm{ran}(\hat{P}_{Su^{\prime}\pm})$ and $\mathrm{ran}(\hat{P}_{Su^{\prime\prime}\pm})$ cannot be elements of one invariant-subspace lattice. So, within the structure of the collection $\{ \mathcal{L}(\Sigma_{Su^{\prime}}), \mathcal{L}(\Sigma_{Su^{\prime\prime}}), \,\dots \}$, the quantum propositions ``Spin of the qubit $S$ along the $u^{\prime\prime}$-axis is $\pm\;\frac{\hbar}{2}\,$'', denoted as $P_{Su^{\prime\prime}\pm}$ and represented by the subspaces $\mathrm{ran}(\hat{P}_{Su^{\prime\prime}\pm})$, are undetermined in any of the states $|\Psi_{Su^{\prime}\pm}\rangle \in \mathrm{ran}(\hat{P}_{Su^{\prime}\pm})$. In symbols,\smallskip

\begin{equation}  
   \Phi_{Su^{\prime\prime}\pm}
   \text{:}
   \;\,
   |\Psi_{Su^{\prime}\pm}\rangle
   \in
   \mathrm{ran}(\hat{P}_{Su^{\prime}\pm})
   \,\cancel{\;\wedge\;}\,
   \mathrm{ran}(\hat{P}_{Su^{\prime\prime}\pm})
   \,
   \not{\!\!\mapsto}
   \;\;
   \mathfrak{t}
   \;\;\;\;  ,
\end{equation}
\smallskip

\noindent where $\Phi_{Su^{\prime\prime}\pm}$ are the predicates for $P_{Su^{\prime\prime}\pm}$.\\

\noindent After the interaction between the qubit $S$ and its environment $E$, the Hilbert space of the composite system $S$E becomes the tensor product\smallskip

\begin{equation}  
   \mathcal{H}_{SE}
   =
   \mathrm{ran}(\hat{1})
   =
   \mathcal{H}_{S}
   \bigotimes_{k=1}^{N}
   \mathcal{H}_{k}
   \;\;\;\;  ,
\end{equation}
\smallskip

\noindent or, explicitly,\smallskip

\begin{equation}  
   \mathcal{H}_{SE}
   =
   \sum_{\beta = \pm}
   \dots
   \left(
      \sum_{\beta = \pm}
      \left(
         \sum_{\beta = \pm}
         \mathcal{H}_{S}
         \otimes
         \mathrm{ran}({\hat{P}}_{1z\beta})
      \right)
      \otimes
      \mathrm{ran}({\hat{P}}_{2z\beta})
   \right)
   \dots
   \otimes
   \mathrm{ran}({\hat{P}}_{Nz\beta})
   \;\;\;\;  .
\end{equation}
\smallskip

\noindent One can observe from this that at some $k$ along the chain $\mathcal{H}_{S} \bigotimes_{k=1}^{N} \mathcal{H}_{k}$ (providing $N$ is large enough) the subspaces $\mathrm{ran}(\hat{P}_{Su^{\prime}\pm})$ and $\mathrm{ran}(\hat{P}_{Su^{\prime\prime}\pm})$ will be factors of tensor products belonging to a common invariant-subspace lattice imposed on the subspaces of $\mathcal{H}_{SE}$.\\

\noindent For the sake of simplicity, assume that this happens at $k = 1$, i.e.,\smallskip

\begin{equation}  
   \mathcal{H}_{SE}
   =
   \mathrm{ran}(\hat{1})
   =
   \left(
      \sum_{\alpha = \pm}
      \mathrm{ran}(\hat{P}_{Su^{\prime}\alpha})
   \!
   \right)
   \!
   \otimes
   \mathrm{ran}(\hat{P}_{1z+})
   +
   \left(
      \sum_{\alpha = \pm}
      \mathrm{ran}(\hat{P}_{Su^{\prime\prime}\alpha})
   \!
   \right)
   \!
   \otimes
   \mathrm{ran}(\hat{P}_{1z-})
   \;\;\;\;  .
\end{equation}
\smallskip

\noindent In the invariant-subspace lattice $\mathcal{L}(\Sigma)$ of the context $\Sigma$ associated with $SE$, namely,\smallskip

\begin{equation}  
   \Sigma
   =
   \left\{
      \hat{P}_{Su^{\prime}+}
      \!
      \otimes
      \hat{P}_{1z-}
      \,
      ,
      \:
      \hat{P}_{Su^{\prime}-}
      \!
      \otimes
      \hat{P}_{1z-}
      \,
      ,
      \:
      \hat{P}_{Su^{\prime\prime}+}
      \!
      \otimes
      \hat{P}_{1z+}
      \,
      ,
      \:
      \hat{P}_{Su^{\prime\prime}-}
      \!
      \otimes
      \hat{P}_{1z+}
   \right\}
   \;\;\;\;  ,
\end{equation}
\smallskip

\noindent the subspaces $\mathrm{ran}(\hat{P}_{Su^{\prime}\pm})$ and $\mathrm{ran}(\hat{P}_{Su^{\prime\prime}\pm})$ are pasted together, that is,\smallskip

\begin{equation}  
   \mathrm{ran}(\hat{P}_{Su^{\prime}\pm})
   \otimes
   \mathrm{ran}(\hat{P}_{1z-})
   \;
   ,
   \:
   \mathrm{ran}(\hat{P}_{Su^{\prime\prime}\pm})
   \otimes
   \mathrm{ran}(\hat{P}_{1z+})
   \:
   \in
   \mathcal{L}(\Sigma)
   \;\;\;\;  .
\end{equation}
\smallskip

\noindent So, the quantum propositions ``Spin of the qubit $S$ along the $u^{\prime\prime}$-axis is $\pm\;\frac{\hbar}{2}$ AND spin of the environmental qubit along the z-axis is $+\;\frac{\hbar}{2}\,$'', denoted as logical conjunctions $P_{Su^{\prime\prime}\pm} \wedge P_{1z+}$ and represented by the tensor products $\mathrm{ran}(\hat{P}_{Su^{\prime\prime}\pm})\otimes\mathrm{ran}(\hat{P}_{1z+})$, are \textit{determined} in the states\smallskip

\begin{equation}  
   |\Psi_{Su^{\prime}\pm}\rangle |\Psi_{1z-}\rangle
   \in
   \mathrm{ran}(\hat{P}_{Su^{\prime}\pm})
   \otimes
   \mathrm{ran}(\hat{P}_{1z-})
   \;\;\;\;  ,
\end{equation}
\smallskip

\noindent where $|a\rangle|b\rangle$ stands, as it is customary, for the tensor product $|a\rangle\!\otimes|b\rangle$. Concretely, since\smallskip

\begin{equation}  
   \mathrm{ran}(\hat{P}_{Su^{\prime}\pm})
   \otimes
   \mathrm{ran}(\hat{P}_{1z-})
   \wedge
   \mathrm{ran}(\hat{P}_{Su^{\prime\prime}\pm})
   \otimes
   \mathrm{ran}(\hat{P}_{1z+})
   =
   \{0\}
   \;\;\;\;  ,
\end{equation}
\smallskip

\noindent the predicates $\Phi_{Su^{\prime\prime}\pm \,\wedge\, 1z+}$ for logical conjunctions $P_{Su^{\prime\prime}\pm} \wedge P_{1z+}$ are:\smallskip

\begin{equation}  
   \Phi_{Su^{\prime\prime}\pm \,\wedge\, 1z+}
   \text{:}
   \;\,
   |\Psi_{Su^{\prime}\pm}\rangle |\Psi_{1z-}\rangle
   \in
   \{0\}
   \;\mapsto\;\,
   \text{false}
   \;\;\;\;  ,
\end{equation}
\smallskip

\noindent which, in accordance with (\ref{PRED}), means that $P_{Su^{\prime\prime}\pm} \wedge P_{1z+}$ assume the value of false in $|\Psi_{Su^{\prime}\pm}\rangle |\Psi_{1z-}\rangle$.\\

\noindent The fact that the environmental qubits have the preferred set of states $\{|\Psi_{kz+}\rangle,|\Psi_{kz-}\rangle\}$ implies that the quantum proposition ``Spin of the environmental qubit along the z-axis is $+\;\frac{\hbar}{2}\,$'', denoted as $P_{1z+}$ and represented by $\mathrm{ran}(\hat{P}_{1z+})$, is false in the composite states $|\Psi_{Su^{\prime}\pm}\rangle |\Psi_{1z-}\rangle$ describing the entanglement between $S$ and $E$.\\

\noindent As $P_{Su^{\prime\prime}\pm} \!\wedge\! P_{1z+}$ and $P_{1z+}$ are definite in these states, it makes sense to say that the quantum propositions $P_{Su^{\prime\prime}+}$ and $P_{Su^{\prime\prime}-}$ have a truth-value there \cite{Foulis, Griffiths}. Besides, since in these states $P_{Su^{\prime\prime}\pm} \wedge P_{1z+}$ and $P_{1z+}$ are all false, the quantum propositions $P_{Su^{\prime\prime}+}$ and $P_{Su^{\prime\prime}-}$ may yield the value of true as well as the value of false in the said states, i.e.,\smallskip

\begin{equation}  
   P_{Su^{\prime\prime}\pm}
   \!
   \left(
      |\Psi_{Su^{\prime}\pm}\rangle |\Psi_{1z-}\rangle
   \right)
   \;\mapsto\;\,
   \{\text{true}, \text{false}\}
   \;\;\;\;  .
\end{equation}
\smallskip

\noindent By reason of this, the result of a collection of experiments intended to determine a truth value of $P_{Su^{\prime\prime}+}$ or $P_{Su^{\prime\prime}-}$ in $|\Psi_{Su^{\prime}\pm}\rangle |\Psi_{1z-}\rangle$ cannot be predicted (based on the knowledge of the state of the system) and is expected to be a distribution describing the numbers of times the values ``true'' and ``false'' are determined.\\

\noindent Considering that the conjunctions $C_{\pm} \!= P_{Su^{\prime\prime}\pm} \!\wedge \neg P_{Su^{\prime\prime}\pm}$ are false in any state and $\neg P_{Su^{\prime\prime}\pm} = P_{Su^{\prime\prime}\mp}$, one finds that the intersection of the events $\mathcal{E}_1 = \{P_{Su^{\prime\prime}+}(|\Psi_{Su^{\prime}\pm}\rangle |\Psi_{1z-}\rangle) \text{ is true}\}$ and $\mathcal{E}_2 = \{P_{Su^{\prime\prime}-}(|\Psi_{Su^{\prime}\pm}\rangle |\Psi_{1z-}\rangle) \text{ is true}\}$, namely, $\mathcal{E}_1 \cap\, \mathcal{E}_2 = \{C_{+}(|\Psi_{Su^{\prime}\pm}\rangle |\Psi_{1z-}\rangle) \text{ is true}\}$, must contain no elements, meaning that $\mathcal{E}_1$ and $\mathcal{E}_2$ must be \textit{mutually exclusive}. This entails the sum rule:\smallskip

\begin{equation}  
   \mathrm{Pr}
   \!
   \left[
      \mathcal{E}_1 \cup \mathcal{E}_2
   \right]
   =
   \mathrm{Pr}[\mathcal{E}_1]
   +
   \mathrm{Pr}[\mathcal{E}_2]
   \;\;\;\;  ,
\end{equation}
\smallskip

\noindent where $\mathrm{Pr}[\cdot]$ denotes the probability of an event. On the other hand, since the disjunction $D = P_{Su^{\prime\prime}+} \!\vee P_{Su^{\prime\prime}-}$ is true in any state, the union of the events $\mathcal{E}_1 \cup \mathcal{E}_2 = \{D(|\Psi_{Su^{\prime}\pm}\rangle |\Psi_{1z-}\rangle) \text{ is true}\}$ must be the entire sample space, and therefore $\mathrm{Pr}[\mathcal{E}_1 \cup \mathcal{E}_2] = 1$.\\

\noindent Given that the states $|\Psi_{Su^{\prime}\pm}\rangle$ can be presented as superpositions of the states $|\Psi_{Su^{\prime\prime}\pm}\rangle \in \mathrm{ran}(\hat{P}_{Su^{\prime\prime}\pm})$ with the coefficients \textit{having equal norms}, namely,\smallskip

\begin{equation}  
   |\Psi_{Su^{\prime}\pm}\rangle
   =
   \frac{1}{\sqrt{2}}
   \left(
      |\Psi_{Su^{\prime\prime}+}\rangle
      \pm
      |\Psi_{Su^{\prime\prime}-}\rangle
   \right)
   \;\;\;\;  ,
\end{equation}
\smallskip

\noindent one can apply \textit{the principle of indifference} and infer that $P_{Su^{\prime\prime}\pm}(|\Psi_{Su^{\prime}\pm}\rangle |\Psi_{1z-}\rangle)$ are \textit{equally likely} to come out true, that is, $\mathrm{Pr}[\mathcal{E}_1] = \mathrm{Pr}[\mathcal{E}_2] = \frac{1}{2}\,$.\\

\section{Conclusion remarks}  

\noindent The main idea underlying probabilistic reasoning about quantum propositions is rather simple.\\

\noindent Let $\{P, Q, \dots\}$ denote the set of all quantum propositions relating to a quantum system and let $b$ stand for \textit{the bivaluation} \cite{Beziau}, i.e., the assignment of truth-values $\mathfrak{t} \in \{\text{true}, \text{false}\}$ to those quantum propositions in the state $|\Psi\rangle$, namely,\smallskip

\begin{equation}  
   b
   \text{:}
   \;\,
   \left\{
      P(|\Psi\rangle)
      ,
      Q(|\Psi\rangle)
      ,
      \dots
   \right\}
   \;\mapsto\;\,
   \{\text{true}, \text{false}\}
   \;\;\;\;  .
\end{equation}
\smallskip

\noindent In this framework, the image of the quantum proposition $P$ in $|\Psi\rangle$ under $b$ is written as $b(P(|\Psi\rangle))$.\\

\noindent Using 1 for true and 0 for false, one can interpret the function $b$ as \textit{the dispersion-free probability measure} $\mu$ such that the probability of $P(|\Psi\rangle)$ being verified is given by\smallskip

\begin{equation}  
   \mathrm{Pr}
   \left[
      P(|\Psi\rangle) \text{ is true}
   \right]
   =
   \mu
   \left(
      P(|\Psi\rangle)
   \right)
   \;\;\;\;  .
\end{equation}
\smallskip

\noindent Drawing the inference from $\mu(P(|\Psi\rangle)) \in \{0,1\}$ one can put $\mu(P(|\Psi\rangle)) \in [0,1]$ as a general principle bringing in this way the logic of quantum propositions to a form of probabilistic reasoning based on the Hilbert space formalism. There is an extensive body of literature demonstrating this approach and its variations, see papers \cite{Sorkin}, \cite{Ghazi}, \cite{Sernadas}, to name but a few.\\

\noindent Even though the derivation of $\mu(P(|\Psi\rangle))$ from $b(P(|\Psi\rangle))$ (according to which $\mu(P(|\Psi\rangle))$ can be regarded as \textit{a probabilistic truth value}) may seem unquestionable from the mathematical point of view, it does not explain how a probability concept appears in the bivaluation of quantum propositions. That is, whence come probabilities in the logic of quantum propositions?\\

\noindent The simple model of the qubit interacting with $N$ other qubits, demonstrated in the present paper, motivates the following answer to this question.\\

\noindent Given that under realistic circumstances the environment $E$ has an enormous number of degrees of freedom, incommutable ranges of projection operators belonging to different contexts of nearly every quantum system $S$ are expected to be pasted together in a common invariant-subspace lattice of a context associated with the composite system $SE$ formed after the inevitable interaction between $S$ and $E$.\\

\noindent On the other hand, according to \textit{the stability criterion} \cite{Schlosshauer}, the environment $E$ (composed of environmental systems) must have the preferred set of states, in which the correlation between any two environmental systems is left undisturbed by the subsequent formation of correlations with other systems of $E$. Invoking this criterion, one finds that in one and the same pure state describing the entanglement between $S$ and $E$, a quantum proposition relating to $S$ may assume the value of true as much as false. Since one cannot predict the truth value of this proposition, one introduces the probability that the proposition in the said state will be verified.\\

\noindent In this sense, the irreducible randomness appearing in the bivaluation of quantum propositions (i.e., randomness which is not related to the uncertainty in the state of the system \cite{Khrennikov}) is \textit{environmentally induced}. Otherwise stated, the interaction between the quantum system and its environment brings on probabilities in the logic of quantum propositions.\\

\bibliographystyle{References}

\end{document}